\documentclass[a4paper]{jpconf}
\usepackage{graphicx}

\begin{document}
\title{First-principles dynamical CPA to finite-temperature magnetism of
transition metals\footnote{to be published in J. Phys.: Conference Series.}}

\author{Y Kakehashi, T Tamashiro, M A R Patoary, and T Nakamura}

\address{University of the Ryukyus, Nishihara, Okinawa, Japan}

\ead{yok@sci.u-ryukyu.ac.jp}

\begin{abstract}
 We present here the first-principles dynamical CPA (coherent potential
 approximation) combined with the tight-binding LMTO LDA+U method towards
 quantitative calculations of the electronic structure and magnetism at
 finite temperatures in transition metals and compounds.  The theory takes 
 into account the single-site dynamical charge and spin
 fluctuations using the functional integral technique as well as an
 effective medium. Numerical results for Fe, Co, and Ni show that the
 theory explains quantitatively the high-temperature properties such as 
 the effective Bohr magneton numbers and the
 excitation spectra in the paramagnetic state, and describes the Curie
 temperatures semiquantitatively. 
\end{abstract}

Quantitative description of the magnetic properties of materials 
has been one of the goals in magnetism.  
For the ground-state properties of transition metals and
alloys, such a calculation scheme has been realized by the density
functional theory (DFT).  The DFT quantitatively explains the 
ground-state magnetizations of Fe, Co, and Ni as well as their cohesive
properties. At finite temperatures, on the other hand, it does not 
explain the local-moment behavior such as the Curie-Weiss
susceptibilities, and overestimates the Curie temperature by a factor of
ten due to neglect of the spin fluctuations.  Because of these
difficulties, quantitative description of the finite temperature magnetism 
has been a long-standing problem in the metallic magnetism \cite{kake04}.

Quite recently, we have proposed the dynamical coherent potential
approximation (CPA) \cite{kake92}
combined with the first-principles tight-binding (TB) linear 
muffin-tin orbital (LMTO) method \cite{ander94} with 
LDA(Local Density Approximation)+U potential \cite{anisimov97}
for band calculations, 
and presented the numerical results for Fe and Ni taking into account
the dynamical corrections up to the second order \cite{kake08}.
In this presentation, we extend the calculations up to the 4th order,
and demonstrate that the theory explains the finite-temperature 
properties of transition metals quantitatively or semiquantitatively.

The dynamical CPA is a dynamical version of the single-site spin
fluctuation theory developed by Cyrot, Hubbard, and Hasegawa
since early in the 1970, and has recently been shown to be equivalent 
to the dynamical mean field
theory (DMFT) \cite{kake04,kake02}.
The theory describes the magnetic properties at finite
temperatures efficiently taking into account the dynamical corrections 
to the static approximation which is exact in the high temperature
limit. Moreover the present approach can treat the transverse spin
fluctuations for arbitrary $d$ electron number, which is different from
the early quantum Monte-Carlo (QMC) calculations combined with 
the DMFT \cite{lich01}.

We adopted the Hamiltonian $H = H_{0} + H_{1}$.  The TB-LMTO Hamiltonian 
$H_{0}$ is written as 
\begin{eqnarray}
H_{0} = \sum_{iL\sigma} \epsilon^{0}_{iL} \, \hat{n}_{iL \sigma} 
+ \sum_{iL jL^{\prime} \sigma} t_{iL jL^{\prime}} \, 
a_{iL \sigma}^{\dagger} a_{jL^{\prime} \sigma} \ ,
\label{h0}
\end{eqnarray}
Here $\epsilon^{0}_{iL}$ is an atomic level on site $i$ and orbital $L$, 
$t_{iL jL^{\prime}}$ is a transfer integral between orbitals $iL$ and 
$jL^{\prime}$. $L=(l,m)$ takes $s$, $p$, and $d$ orbitals. 
$a_{iL \sigma}^{\dagger}$ 
($a_{iL \sigma}$) is the creation (annihilation) operator for an
electron with orbital $L$ and spin $\sigma$ on site $i$, and 
$\hat{n}_{iL\sigma}=a_{iL \sigma}^{\dagger}a_{iL \sigma}$ is a charge
density operator for electrons with orbital $L$ and spin $\sigma$ on
site $i$.  
It should be noted that $\epsilon^{0}_{iL}$ is not identical with the
atomic level $\epsilon_{iL}$ in the DFT-LDA potential because the latter 
contains exchange correlation contributions.
For the atomic $d$ level, we adopted the formula 
$\epsilon^{0}_{iL} = \epsilon_{iL} - \partial
E^{U}_{\rm LDA}/\partial n_{iL\sigma}$ \cite{anisimov97}. 
Here $n_{iL\sigma}$ is the
charge density at the ground state, $E^{U}_{\rm LDA}$ is a LDA
functional to the intraatomic Coulomb interactions.
We adopted the Hartree-Fock type form for $E^{U}_{\rm LDA}$ proposed by 
Anisimov {\it et al.} \cite{anisimov93} 
because we treat here the itinerant electron
system where the ratio of the Coulomb interaction to the $d$ band width
is not larger than one. For the other orbital electrons, we did not 
make any corrections for simplicity.

The intraatomic Coulomb interactions $H_{1}$ is given as  
\begin{eqnarray}
H_{1} = \sum_{i} 
\Big[ \sum_{m} U_{0} \, \hat{n}_{ilm \uparrow} \hat{n}_{ilm \downarrow} 
+ {\sum_{m > m^{\prime}}} 
(U_{1}-\frac{1}{2}J) \hat{n}_{ilm} \hat{n}_{ilm^{\prime}} -
{\sum_{m > m^{\prime}}} J   
\hat{\mbox{\boldmath$s$}}_{ilm} \cdot \hat{\mbox{\boldmath$s$}}_{ilm^{\prime}} 
\Big] \ . 
\label{h1}
\end{eqnarray}
Here $U_{0}$ ($U_{1}$) and $J$ in the interaction $H_{1}$ are 
the intra-orbital (inter-orbital)
Coulomb interaction and the exchange interaction, respectively.  
$\hat{n}_{ilm}$ ($\hat{\mbox{\boldmath$s$}}_{ilm}$) with $l=2$ is 
the charge (spin)
density operator for $d$ electrons on site $i$ and orbital $m$.

We have presented the first-principles dynamical CPA in our previous 
paper \cite{kake08}.
We briefly explain here the outline. In the theory, 
we transform the interacting Hamiltonian $H_{1}$
into a dynamical potential $v$ in the free energy adopting the
functional integral method.  Introducing a uniform medium, ({\it i.e.} 
a coherent potential) $\Sigma_{L\sigma}(i\omega_{n})$, 
we expand the free energy with respect
to sites.  Note that $\omega_{n}$ denotes the Matsubara
frequency.  The first term in the expansion is the free energy for 
a uniform medium, 
$\tilde{\cal F}[\mbox{\boldmath$\Sigma$}]$.  
The second term is an impurity contribution to the
free energy.  The dynamical CPA neglects the higher-order terms, so that
the free energy per atom is given by
\begin{eqnarray}
{\mathcal F}_{\rm CPA} = \tilde{\mathcal F}[\mbox{\boldmath$\Sigma$}]
- \beta^{-1} {\rm ln} \, C \int d\mbox{\boldmath$\xi$} \,
{\rm e}^{\displaystyle -\beta E_{\rm eff}(\mbox{\boldmath$\xi$})} .
\label{fcpa2}
\end{eqnarray}
Here $\beta$ is the inverse temperature.  $C$ is a normalization
constant.  $\mbox{\boldmath$\xi$}$ is the static field variable on a site.  
$E_{\rm eff}(\mbox{\boldmath$\xi$})$ is an effective potential 
projected onto the static field $\mbox{\boldmath$\xi$}$.  
It consists of the static
contribution $E_{\rm st}(\mbox{\boldmath$\xi$})$ and the dynamical
correction term 
$E_{\rm dyn}(\mbox{\boldmath$\xi$})$.  
We obtain the latter using the harmonic approximation 
(see Ref. \cite{kake08} for details of expressions of 
$E_{\rm st}(\mbox{\boldmath$\xi$})$ and 
$E_{\rm dyn}(\mbox{\boldmath$\xi$})$).   

The effective medium, {\it i.e.}, the coherent potential can be
determined by the stationary condition
$\delta \mathcal{F}_{\rm CPA}/\delta \Sigma = 0$.  
This yields the CPA equation as 
\begin{eqnarray}
\langle G_{L\sigma}(\mbox{\boldmath$\xi$}, i\omega_{l}) \rangle 
= F_{L\sigma}(i\omega_{l}) \ .
\label{dcpa3}
\end{eqnarray}
Note that $\langle \ \rangle$ at the r.h.s. 
is a classical average taken with respect to the
effective potential $E_{\rm eff}(\mbox{\boldmath$\xi$})$, 
$ F_{L\sigma}(i\omega_{l}) = [(i\omega_{l} - \mbox{\boldmath$H$}_{0} 
- \mbox{\boldmath$\Sigma$})^{-1}]_{iL\sigma iL\sigma}$ 
is the coherent Green function, where 
$(\mbox{\boldmath$H$}_{0})_{iL\sigma jL^{\prime}\sigma}$ is the 
one-electron TB-LMTO Hamiltonian matrix, and 
$(\mbox{\boldmath$\Sigma$})_{iL\sigma jL^{\prime}\sigma} = 
\Sigma_{L\sigma}(i\omega_{l})\delta_{ij}\delta_{LL^{\prime}}$.
Furthermore, $G_{L\sigma}(\mbox{\boldmath$\xi$}, i\omega_{l})$
is the impurity Green function given by
\begin{eqnarray}
G_{L\sigma}(\mbox{\boldmath$\xi$}, i\omega_{l}) = 
\tilde{g}_{L\sigma\sigma}(i\omega_{l}) - 
\frac{\displaystyle \beta}{\displaystyle \kappa_{L\sigma}(i\omega_{l})}
\frac{\displaystyle \delta E_{\rm dyn}(\mbox{\boldmath$\xi$})}
{\displaystyle \delta \Sigma_{L\sigma}(i\omega_{l})} \ .
\label{gimp}
\end{eqnarray}
Here the first term at the r.h.s. is the Green function in the static
approximation. 
The second term is the dynamical corrections, and 
$\kappa_{L\sigma}(i\omega_{l})= 1 - F_{L\sigma}(i\omega_{l})^{-2}
\delta F_{L\sigma}(i\omega_{l})/\delta \Sigma_{L\sigma}(i\omega_{l})$.

Solving the CPA equation (4) self-consistently, we obtain the effective
medium.  The magnetic moment is then given by 
$\langle \hat{m}^{z}_{L} \rangle = 
\beta^{-1} \sum_{n\sigma} \sigma F_{L\sigma}(i\omega_{n})$,
and the density of states (DOS) 
is obtained by means of a numerical analytic continuation. 

In the numerical calculations, 
the average Coulomb and exchange energy parameters 
($\overline{U}$ and $J$) are
taken from the LDA+U band calculations \cite{anisimov97,banddyo89};
$(\overline{U},J)= (0.169,0.066), (0.245, 0.069), (0.221, 0.066)$
Ry for Fe, Co, and Ni, respectively.
$U_{0}$ and $U_{1}$ are obtained from the relations 
$U_{0} = \overline{U}+8J/5$ and $U_{1} = \overline{U} - 2J/5$.
These sets of parameters yield the Hartree-Fock Curie temperatures:
12200K, 12100K, and 4950K for Fe, Co, and Ni, respectively.
\begin{figure}[t]
\includegraphics[width=18pc]{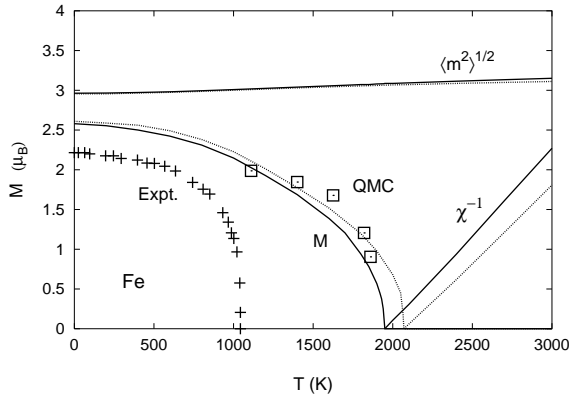}\hspace{2pc}%
\begin{minipage}[b]{18pc}\caption{\label{femt}
Magnetization ($M$), inverse susceptibility ($\chi^{-1}$), and amplitude
 of local moment ($\langle {\bf m}^{2} \rangle^{1/2}$) 
vs. temperature ($T$) curves for Fe.
The dynamical results are shown by the solid curves, while 
the results in the static approximation are shown by dotted curves.
The DMFT results without transverse spin fluctuations~\cite{lich01} 
are shown by open squares.
Experimental data of magnetization~\cite{potter34} are shown by $+$. 
}
\end{minipage}
\end{figure}

We present in Fig. 1 the results of calculations for Fe.
The calculated susceptibility follows the Curie-Weiss law with the
effective Bohr magneton number $m_{\rm eff}=3.0 \mu_{\rm B}$ in
agreement with the experimental value $3.2 \mu_{\rm B}$.
We find the Curie temperature $T_{\rm C}=2070$K in the static
approximation.  By taking into account the dynamical corrections, the
Curie temperature reduces to 1930 K, which is comparable to 1900K
obtained by the QMC calculations of the DMFT without
transverse exchange interaction.  The theoretical $T_{\rm C}$ is 
higher than the experimental result 1043K 
by a factor of 1.8.  The ground-state 
magnetization obtained by an extrapolation, 2.58 $\mu_{\rm B}$ is also
considerably overestimated as compared with the experimental value 2.22
$\mu_{\rm B}$.  

In the case of Ni, we obtained the Curie temperature
$T_{\rm C}=1420$K in the static approximation as shown in Fig. 2.  
The dynamical corrections much reduce the Curie temperature; 
$T_{\rm C}=620$K, which agrees well with the
experimental value 630K.  Calculated inverse susceptibility shows
slightly upward convexity.  We obtained the effective Bohr magneton
number $m_{\rm eff}=1.6 \mu_{\rm B}$ in the high temperature region. 
The result agrees with the experimental one $1.6 \mu_{\rm B}$.  
As seen from Fig. 3, a large
reduction of $T_{\rm C}$ in the dynamical CPA calculation 
seems to be related to the
reduction of the DOS $\rho (0)$ at the Fermi level when we take into
account the third and fourth order dynamical corrections.  The behavior is
consistent with the argument of the stability of ferromagnetism at the
ground state \cite{kana63}.
\begin{figure}[h]
\begin{minipage}{18pc}
\includegraphics[width=18pc]{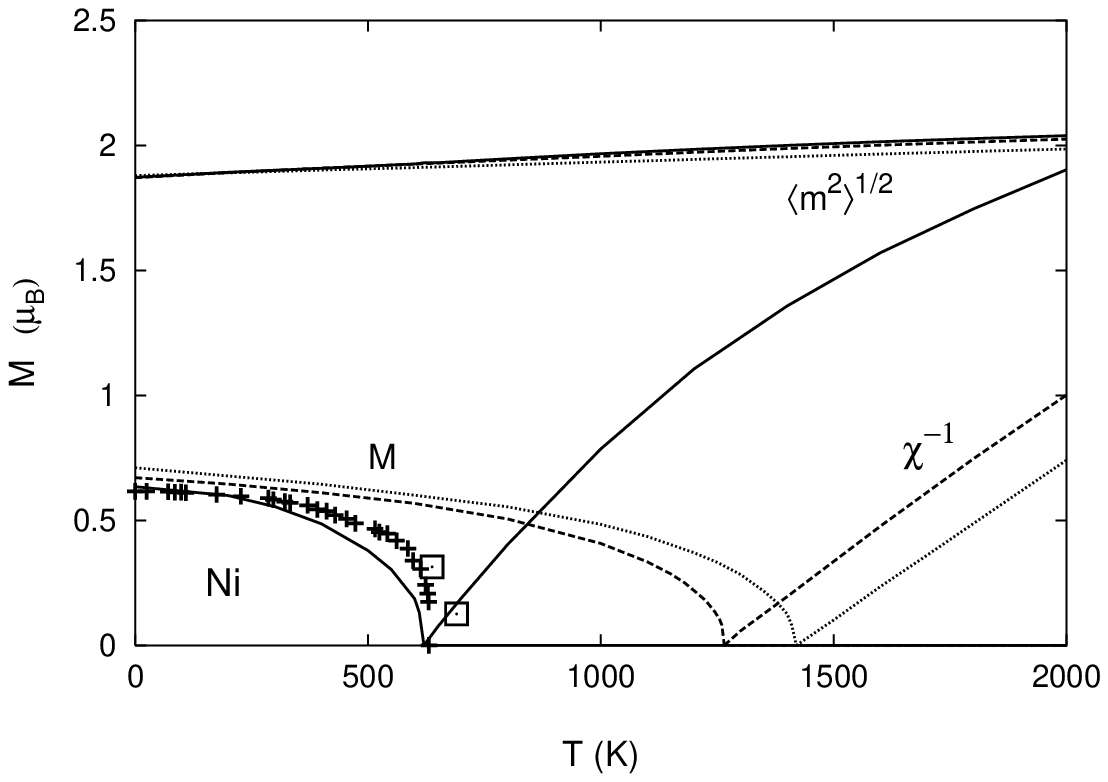}
\caption{\label{nimt}
Magnetization, inverse susceptibility, and amplitude of local moment as
 a function of temperature for Ni.
The dynamical results are shown by the solid curves, while the results 
in the static calculation are shown by dotted curves.
Dashed curves show the results of the 2nd-order dynamical CPA.
The DMFT results~\cite{lich01} are shown by open squares,
and experimental data of magnetization~\cite{weiss26} are shown by $+$. 
}
\end{minipage}\hspace{2pc}%
\begin{minipage}{18pc}
\includegraphics[width=18pc]{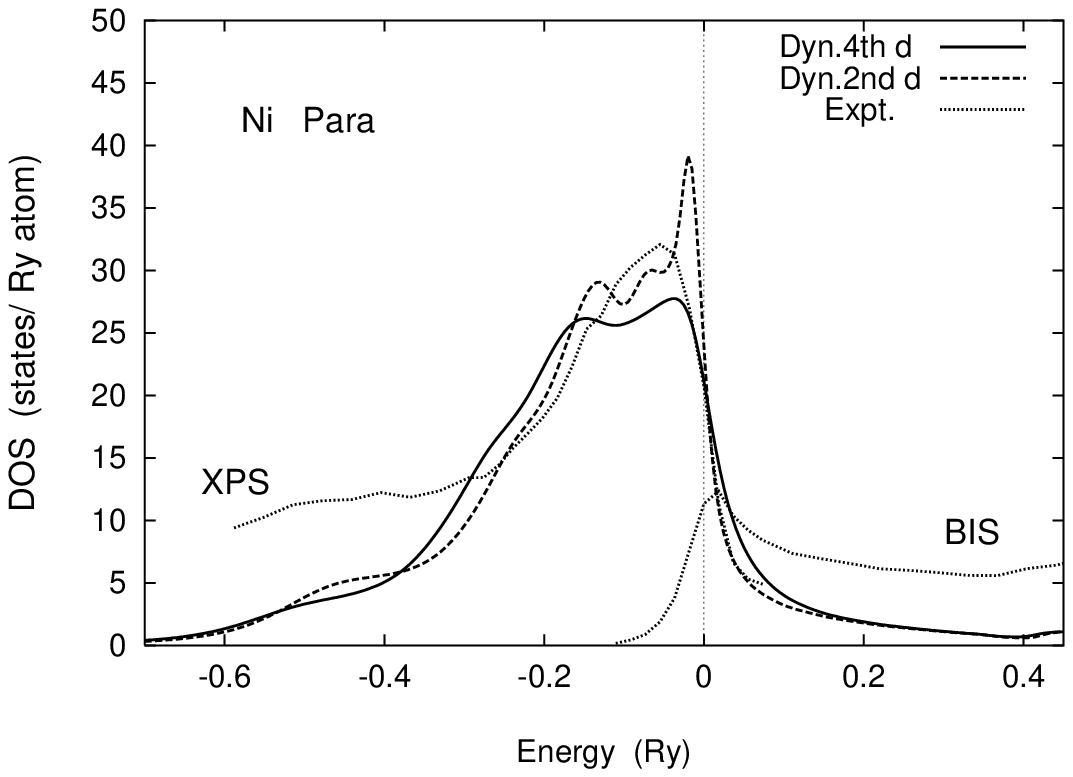}
\caption{\label{nidos}Calculated DOS in the paramagnetic Ni.
Solid curve: dynamical CPA, dashed curve:
 2nd-order dynamical CPA, dotted curves: XPS and BIS experimental 
data~\cite{narmo88,speier84}.  
}
\end{minipage} 
\end{figure}

We have also investigated the finite temperature magnetism for the fcc
Co.  In this case, we obtained $T_{\rm C}=2550$K which is larger
than the experimental value 1388K by a factor of 1.8.  Calculated
effective Bohr magneton number of Co is $3.0 \mu_{\rm B}$, being in good
agreement with the experimental value $3.15 \mu_{\rm B}$.
We have also calculated the densities of states from Sc to Cu
in the high-temperature region, 
where the present theory works best, and found that the DOS explain 
well the XPS and BIS data.  The correlation effects on these 
excitation spectra will be discussed in a separate paper.

In summary, we have presented the dynamical CPA combined with the
first-principles TB-LMTO LDA+U Hamiltonian.
Calculated Curie temperature $T_{\rm C}$ for Ni agrees with the
experiment, but $T_{\rm C}$ for Fe and Co are higher than the
experimental ones by a factor of 1.8.  On the
other hand, the present theory explains quantitatively the effective Bohr
magneton numbers as well as the excitation spectra in the high
temperature region.  In order to obtain observed $T_{\rm C}$, we have to 
take into account the nonlocal correlations definitely.  Such an attempt 
is in progress.  

This work was supported by Grant-in-Aid for Scientific Research (19540408). 
Numerical calculations were carried out with use of the Hitachi SR11000 
in the Supercomputer Center, Institute of Solid State Physics, 
University of Tokyo.

\end{document}